\documentclass[twocolumn,prd]{revtex4}

\usepackage[dvips]{graphics, color}
\usepackage{alltt}

\begin{document}
\title{A Statistically Robust 3-$\sigma$ Detection of Non-Gaussianity 
in the WMAP Data Using Hot and Cold Spots}
\author{David L. Larson} 
\email{dlarson1@uiuc.edu}
\affiliation{Department of Physics, University of Illinois at 
Urbana-Champaign, Champaign, IL 61801}
\author{Benjamin D. Wandelt}
\affiliation{Department of Astronomy, University of Illinois at 
Urbana-Champaign, Champaign, IL 61801}
\affiliation{Department of Physics, University of Illinois at 
Urbana-Champaign, Champaign, IL 61801}
\date{\today}

\keywords{Cosmology, CMB, Cosmic Microwave Background, WMAP, Gaussianity, Statistical test}
\pacs{98.80.Jk, 98.80.Bp, 98.80.Es, 98.80.-k}

\begin{abstract}
We present a careful frequentist analysis of 
one- and two-point statistics of the hot and cold spots in the cosmic
microwave background (CMB) data obtained by the Wilkinson Microwave Anisotropy
Probe (WMAP). Our main result is the detection of a
new anomaly at the 3-sigma level using temperature-weighted extrema
correlation functions.  We obtain this result using a simple hypothesis test
which reduces the maximum risk of a false detection to the same
level as the claimed significance of the test. We further present a detailed 
study of the robustness
of our earlier claim (Larson and Wandelt 2004) under variations in the noise
model and in the resolution of the map.  Free software which implements our
test is available online.
\end{abstract}

\maketitle

\section{Introduction}

Inflation  predicts Gaussian random density 
fluctuations in the early Universe.  This prediction can be tested by observing
the Cosmic Microwave Background (CMB) anisotropies.  
By observing the CMB, we look back early
enough to check if these initial conditions of the Universe
were laid down by inflation.
Density fluctuations produced in standard models of inflation will cause the
CMB to be a highly Gaussian isotropic random field --- a directly testable 
prediction
since the high-resolution, all-sky data set of the Wilkinson Microwave
Anisotropy Probe (WMAP) has become available \cite{BennettHalpern03}.

In this paper we check that prediction, continuing our work in 
\cite{LarsonWandelt04}, hereafter LW04.
As in that paper, we look at the pattern of hot and cold spots (local extrema) 
seen in the CMB sky. 
We use several methods to determine if that pattern is statistically
similar to the patterns of hot and cold spots that we simulate for Gaussian
isotropic random fields on
the sky.  In LW04, we described an anomaly of the one-point statistics of hot
and cold spot excursions.
In this paper, we extend this earlier analysis and add a detailed study of 
two-point
statistics, the spot-spot and temperature-weighted correlation functions of
hot and cold spots. 

Most frequentist searches for non-Gaussianity follow a set recipe: 
first compare a statistic computed on
observed data to a set of Monte Carlo simulations. An assessment of
goodness-of-fit  then leads to a  significance level at which Gaussianity is
rejected. This assessment is often rather informal and prone to false
detection. One of the main goals of 
our paper is the presentation of a robust hypothesis test that is applicable
to all tests of Gaussianity in  this category.

A number of groups have applied this recipe to  various
statistics of the CMB anisotropy.
For example, Vielva et~al.\ detect non-Gaussianity in the three- and 
four-point wavelet moments
\cite{VielvaMartinezGonzalez03}, 
as do Liu and Zhang \cite{LiuZhang05}, who claim it may be a residual 
foreground.
McEwen et al.~investigate wavelets that aren't azimuthally symmetric, and 
find non-Gaussianity 
using the skewness and kurtosis of their wavelet coefficients 
\cite{McEwenHobson05}.
Chiang et~al.\ detect non-Gaussianity in phase correlations
between spherical harmonic coefficients 
\cite{ChiangNaselsky03, ChiangColes02, ChiangNaselsky04},
and Park finds it in the genus Minkowski functional \cite{Park04}. 
Eriksen et~al.\ find anisotropy in the $n$-point
functions of the CMB in different patches of the sky \cite{EriksenHansen04}.
Others discuss possible methods of detecting non-Gaussianity.
Aliaga et~al.\ look at studying non-Gaussianity through spherical wavelets and ``smooth tests of
goodness-of-fit'' \cite{AliagaMartinezGonzalez03}. 
Cabella et~al.\ review three methods of studying non-Gaussianity: through Minkowski functionals,
spherical wavelets, and the spherical harmonics \cite{CabellaHansen04}. 
They propose a way to combine these methods.
More recently, Cabella et al.~constrain one generic type of non-Gaussianity using spherical wavelets and local curvature of
the CMB temperature field \cite{CabellaLiguori05}.
Komatsu et~al.\ discuss a fast way to test the bispectrum for primordial non-Gaussianity in the CMB  
\cite{KomatsuKogut03}, 
and do not detect it \cite{KomatsuSpergel03}. 
Using a generic model for non-Gaussianity, Babich shows that the bispectrum
is the best way to constrain it, and therefore claims that the bispectrum test used
by the WMAP team (Komatsu et al.) is optimal \cite{Babich05}.
Finally, Gazta{\~n}aga et al.\ find the CMB to be consistent with Gaussianity when considering the
two and three-point functions 
\cite{GaztanagaWagg03a, GaztanagaWagg03b}.

This paper is laid out as follows.  In section 2, we briefly discuss the
statistical approach underlying all frequentist blind searches of
non-Gaussianity.  In section 3, we study the one-point 
statistics of hot and cold spots at several map resolutions and for various
noise models.  In
Section 4 we remove dependence on the noise model by smoothing and describe
the spot-spot and temperature 
weighted correlation functions of hot and cold spots at 50' and 3 degrees. We
conclude in section 5. The Appendix  describes our robust hypothesis test,
as well as the free software which implements it.

\section{Statistical Analysis}

The problem we tackle in this paper is how to find non-Gaussianity in the CMB
when we do not have a specific model for the non-Gaussianity.  We have a measured CMB sky (WMAP data)
and the ability to make many Gaussian simulations of CMB skies, and we want to determine
if the measured CMB sky looks like it came from the distribution of skies we simulate.

Because we do not have a model for non-Gaussianity, we do not have a second distribution
of skies to compare to the Gaussian distribution.  This limits the testing we can do to
merely determining how large a statistical fluctuation our currently measured CMB sky is.

We approach the problem numerically as follows: we find some way to reduce the an entire CMB sky to 
a single number, a single statistic computed on the hot and cold spots.  
We compare the statistic for the measured CMB sky to the distribution of statistics for the simulated CMB skies.  
If the measured statistic falls significantly higher or lower than all of the others, 
then we have a large statistical fluctuation, which we quantify.  It is then up to the reader
to determine if this should be interpreted as merely an unlikely statistical fluctuation, an indication
of non-Gaussianity, a residual foreground, or a mismatch between  our
simulations and the actual observations. 
To eliminate that last possibility, we describe our simulations in detail.

The specific simulation methods and statistics we use are detailed in the corresponding sections.  
A detailed discussion of what constitutes a statistically significant detection is
given in 
the Appendix and accompanying freely distributed \emph{Mathematica} notebook and C code.

\section{Multi-resolution One-point Analysis}

\subsection{WMAP Measurement}

We construct a single temperature map of the CMB to represent
the WMAP team's measurements.
Motivated by our multi-frequency study in LW04, 
we take this map to be an unweighted average of the four channels of the 
W-band foreground cleaned temperature maps
(available on the LAMBDA web site\footnote{http://lambda.gsfc.nasa.gov/}.  
We use an \emph{unweighted} average of the temperature maps so that we can take the (azimuthally symmetric) beam
to simply be the average of the beams for each of the channels.
This is the map whose properties we concentrate on simulating.

\subsection{Simulation Process}
\label{noiseScale}

We take legitimate shortcuts in our Gaussian sky simulation process.
The brute-force way to simulate the WMAP team's measurement of a Gaussian CMB sky
is to simulate the CMB sky, add foregrounds, simulate the time-ordered-data stream
from the WMAP satellite, and then run it through the full map-making and 
foreground removal pipeline.  We take a simpler approach; we simulate the
result of that process by
a Gaussian CMB sky and then add either white or correlated noise to that sky.
This is acceptable because the full WMAP pipeline does a good job of reconstructing
some characteristics of the true CMB sky, and we are careful to make sure
our statistics depend only on those characteristics.

One possible criticism of LW04 is that we used uncorrelated noise in our CMB simulations.  
There are good reasons for using uncorrelated noise: the noise is
stated to be white noise, and the primary correlations are at angular scales of about 141 degrees 
and are on the order of 0.3\% \cite{HinshawBarnes03}.  
For this paper, however, we compare the results of white and correlated noise.

The WMAP team has provided 110 publicly available \cite{LAMBDA} correlated noise simulations 
that can be used to calculate the correlated noise on our averaged W-band map.
We take each of the simulated noise maps provided by the WMAP team and
average and rescale them.  
Since there were 4 channels in the W band, we do an unweighted average of the noise maps.   
Then we rescale the noise, pixel-by-pixel, so that the number of effective observations
(amount of noise in that pixel) exactly matches the amount quoted in the measured W band.
This is necessary because the number of observations in the simulated data provided by the WMAP team does not exactly
match the number of observations in the measured data \cite{HinshawEmail}.
This requires about a 2.5\% decrease in the noise.  Since the correction is approximately the
same in each pixel, this rescaling should maintain the correlations in the noise.

Our white noise simulations model the noise as an unweighted average of Gaussian noise
in each of the four W-band channels.

\subsection{Data Reduction}

Care must be taken in our reduction of the data to a single statistic.
Specifically, the reduction process should ignore data contaminated by 
galactic foregrounds, and should be insensitive to monopole and dipole moments.
In this section we describe our data reduction process:
lowering the resolution of the HEALPix\footnote{http://www.eso.org/science/healpix/} map,
ignoring data contaminated by Galactic foregrounds, 
removing the monopole and dipole moments,
finding the local maxima and minima in temperature, 
and calculating statistics on these local extrema.

Lowering the resolution allows us to investigate the dependence of our results on
different angular scales.  Specifically, we can see how the white
noise and correlated noise models behave at different resolutions.
In this case, lowering the resolution simply means doing an unweighted average
of all the smaller pixels inside the larger degraded pixel.

\begin{figure*}
\resizebox{12cm}{!}{
	\includegraphics{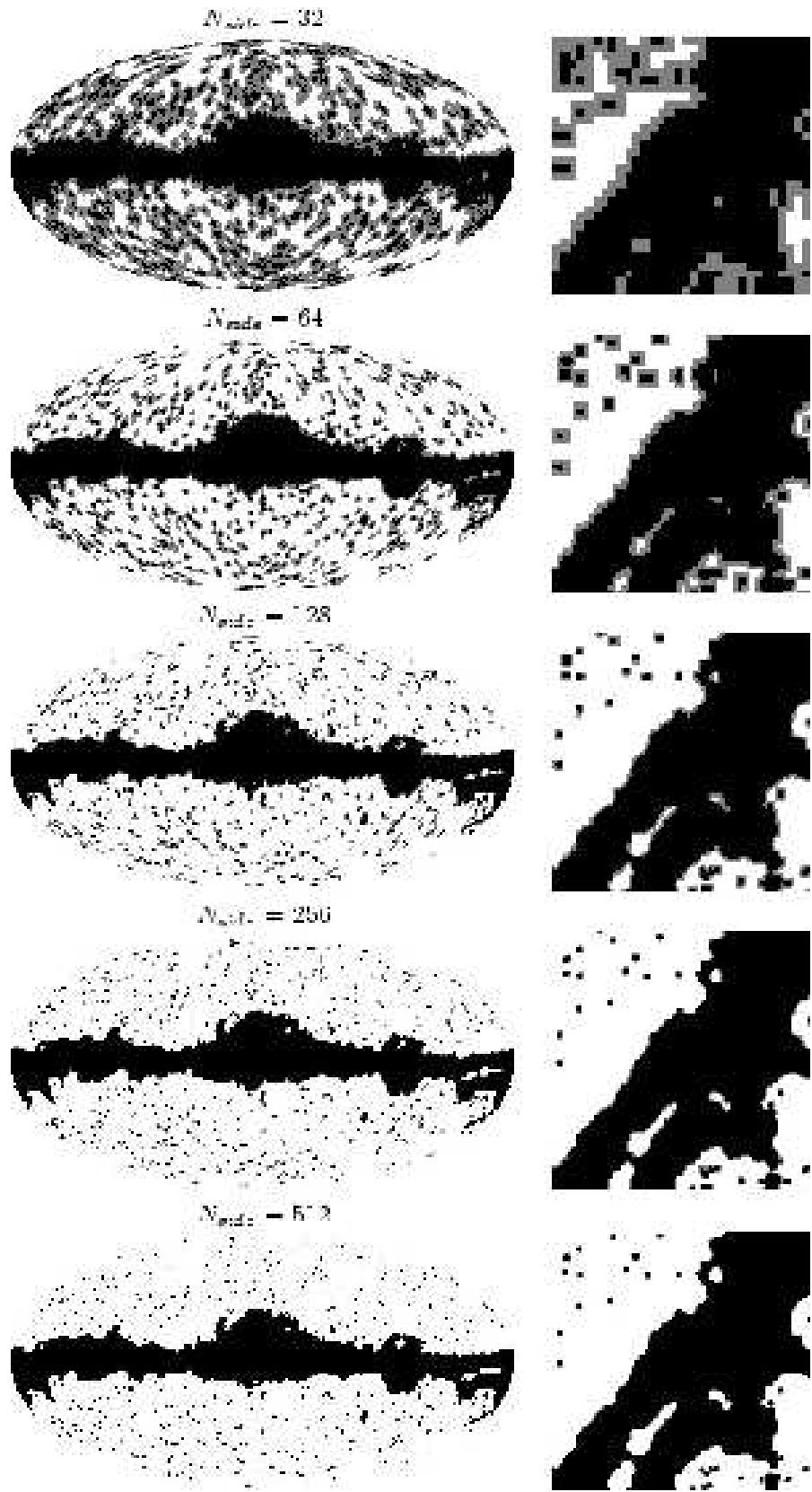}
}
\caption{
\label{onePointMasks}
These are the 5 masks used for the one-point correlations.  
Left column is mollweide projection of the sky; right column is HEALPix base tile 6, 
where the upper left corner is northernmost.  Tile 6 is directly opposite the galactic center;
it's solid angle is exactly 1/12 of the full sphere's.
Paranoid mask is black, extended paranoid mask extends it in grey.  
}
\end{figure*}

We use several masks to be sure we have removed all the effects of the Galactic foregrounds.
See Figure \ref{onePointMasks}.
We start with the kp0 mask, which we must degrade to a lower resolution.  Every degraded
(larger) pixel which contains any of the original kp0 mask is also masked; we call
this degraded mask the ``paranoid'' mask.  We extend this mask by one pixel in all directions
to get a ``paranoid extended'' mask, which will be useful later.
We do not apply our analysis to different hemispheres for our one-point statistics, because at low resolution
the degraded masks would block too much of the sky.

We remove the monopole and dipole moments from the map outside the paranoid mask, and then find the local extrema.  
Since these extrema are defined by their neighboring
pixel values, extrema next to the paranoid mask are dependent on pixels we wish to mask.
To be completely independent of masked pixels, we
ignore all extrema inside the paranoid extended mask.

Finally, we calculate statistics on the local extrema as in LW04.
We use a one-point analysis involving statistics that ignore their angular distribution. 
For the hot spots, we use the
number of hot spots, as well as the mean temperature, and variance, skewness and kurtosis of the 
temperatures.  We calculate the same statistics for the cold spots.
For completeness, we give the formulas here, where angle brackets represent an average over
all hot (or cold) spot temperatures $t$:
\begin{eqnarray}
\mbox{mean} & = & \langle t \rangle  \nonumber\\ 
\mbox{variance} & = & \langle(t - \langle t\rangle)^2 \rangle \nonumber\\ 
\mbox{skewness} & = & \frac{\langle(t - \langle t\rangle)^3 \rangle}{\mbox{variance}^{3/2}} \nonumber\\ 
\mbox{kurtosis} & = & \frac{\langle(t - \langle t\rangle)^4 \rangle}{\mbox{variance}^2} 
\end{eqnarray}

The process of reducing the data to a single statistic is summarized by the following list.
We use identical methods to reduce both the WMAP data and the simulations.

\begin{enumerate}
\item Degrade the map resolution the desired resolution.
\item Remove the dipole from the map outside of the paranoid mask.
\item Find the local extrema.
\item Ignore local extrema inside the paranoid extended mask.
\item Calculate statistics of the  remaining local extrema as usual.
\end{enumerate}

\subsection{One-Point Multi-Resolution Results}

\begin{figure*}
\resizebox{17cm}{!}{
	\includegraphics{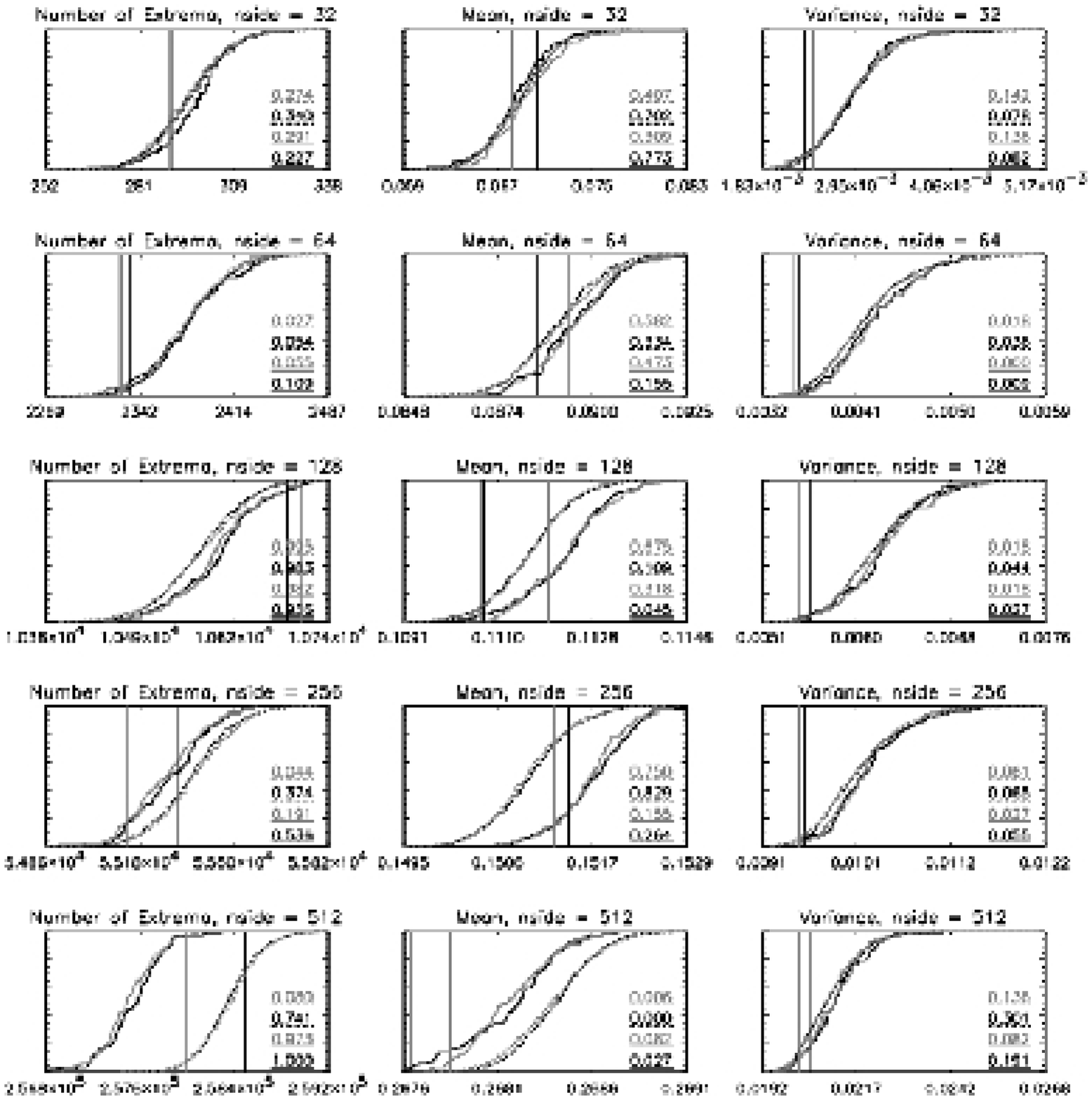}
}
\caption{
\label{combinedOnePointCDFs}
Cumulative distribution functions of the hot and cold spot statistics from various simulations.
The grey (black) dashed line is the CDF for the hot (cold) spots
from the 800 white noise simulations.  The grey (black) solid line is the hot (cold) spot CDF
from the 110 correlated noise simulations.  The vertical grey (black) line gives 
the location of the WMAP statistic for the hot (cold) spots.  Estimates $\hat{p}$ of the probability
of a simulation's statistic falling below the WMAP statistic are printed on the graph and
underlined with the appropriate line.
All simulations use the best fit power spectrum, the kp0 mask (degrading is detailed in the paper), 
and data from the W band.  Units for the mean statistics are $mK$ and units for the variance are $mK^2$.
}
\end{figure*}

From the cumulative distribution functions (CDFs) in figure 
\ref{combinedOnePointCDFs}, one can see how the difference between 
correlated noise and white noise changes with resolution.  The resolution has a dramatic effect
on the number of extrema and on their mean value.  At lower resolutions, the CDFs for the white
noise and correlated noise look very similar, as one would expect.

It is true that switching from white noise to correlated noise reduces our original detection
of the hot and cold spots not being hot and cold enough.   However, the switch to
correlated noise reduces the number of hot and cold spots in the simulations, so now we have 
too many extrema.  Whether or not there are too many extrema also seems to be heavily dependent on 
the resolution at which we find the extrema.  

At all scales, we do find the variance of the extrema to be slightly low.

Because we have only 110 correlated noise samples, we cannot claim any 95\% fluctuations for 
this noise model.  We can claim occasional 95\% fluctuations for the 800 white noise samples we have,
but in light of the dramatic differences between the CDFs, unlikely statistics from the white noise simulations
are more likely to be indications of an incorrect noise model than of non-Gaussianity.  
Nonetheless, we do calculate where 95\% fluctuations occur and mark them in figure \ref{onePointPHats}.

\begin{figure*}
\begin{center}
\begin{tabular}{lllllllllll}
\hline
 & \multicolumn{2}{c}{Number}
 & \multicolumn{2}{c}{Mean}
 & \multicolumn{2}{c}{Variance}
 & \multicolumn{2}{c}{Skewness}
 & \multicolumn{2}{c}{Kurtosis}\\
 & white & corr.
 & white & corr.
 & white & corr.
 & white & corr.
 & white & corr.
 \\
\hline
 32, max & 0.274 & 0.291 & 0.407 & 0.309 & 0.142 & 0.136 & 0.391 & 0.427 & 0.114 & 0.118\\
 32, min & 0.340 & 0.227 & 0.702 & 0.773 & 0.078 & 0.082 & 0.928 & 0.927 & 0.161 & 0.136\\
 64, max & 0.027 & 0.055 & 0.582 & 0.473 & 0.018 & 0.000 & 0.390 & 0.482 & 0.940 & 0.964\\
 64, min & 0.054 & 0.109 & 0.334 & 0.155 & 0.036 & 0.009 & 0.887 & 0.827 & 0.441 & 0.491\\
128, max & 0.995*  & 0.982 & 0.675 & 0.318 & 0.018 & 0.018 & 0.141 & 0.100 & 0.990*  & 0.973\\
128, min & 0.983 & 0.936 & 0.109 & 0.045 & 0.044 & 0.027 & 0.794 & 0.800 & 0.950 & 0.936\\
256, max & 0.044 & 0.191 & 0.750 & 0.155 & 0.061 & 0.027 & 0.610 & 0.591 & 0.900 & 0.900\\
256, min & 0.374 & 0.536 & 0.829 & 0.264 & 0.065 & 0.055 & 0.822 & 0.745 & 0.683 & 0.709\\
512, max & 0.080 & 0.973 & 0.006*  & 0.082 & 0.136 & 0.082 & 0.426 & 0.527 & 0.385 & 0.445\\
512, min & 0.741 & 1.000 & 0.000*  & 0.027 & 0.301 & 0.191 & 0.853 & 0.845 & 0.627 & 0.682\\
\hline
\end{tabular}
\end{center}
\caption{
\label{onePointPHats}
Estimates $\hat{p}$ of the probability $p$ that a simulation statistic will be lower
than the WMAP statistic.  Column headings signify the statistic and whether the noise
was white or correlated in the simulations.  For this table, we used 110 correlated
noise simulations and 800 white noise simulations.  Rows labels signify the value
of $N_{side}$ and whether the statistics are for maxima or minima.  Values of $\hat{p}$ 
that are significant for our 95\% level test have asterisks.  Only the white noise
has enough simulations to enable a 95\% detection.  The first 6 columns of data are presented graphically
in figure \ref{combinedOnePointCDFs}.
}
\end{figure*}

\subsection{Varying the Amplitude of the Noise}

\begin{figure*}
\resizebox{17cm}{!}{
	\includegraphics{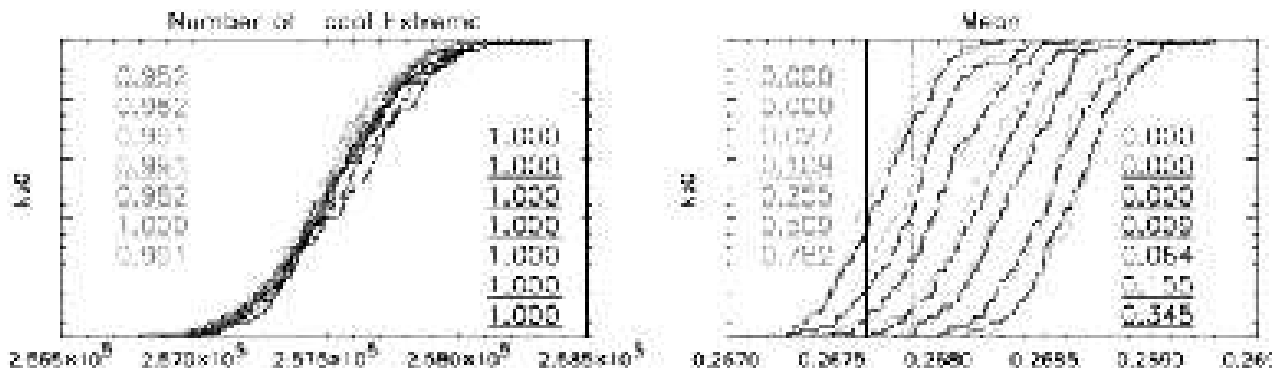}
}
\caption{
\label{amplitudeShiftFig}
This figure shows the cumulative distribution functions for the statistics
generated by shifting the correlated noise amplitude by $n\sigma$, where $n$ ranges from -3 to 3.
110 correlated noise maps are used.
The grey statistics (and dotted lines) are for maxima, the black (and solid lines) for minima.  
From top to bottom, the numbers are values of $\hat{p}$ for $n=3$ to $n=-3$.  
The vertical lines are the WMAP values.  
Note: Only the Mean statistic (plot on the right) shows the CDFs spread apart from each other; 
the other 4 statistics
have the CDFs on top of each other, as in the left plot.  Also, the detection of non-Gaussianity
in the mean is weaker when you consider a $-1\sigma$ shift in the correlated noise amplitude.
The CDFs would become smoother with more than 110 simulations.
}
\end{figure*}

We also check the effect of varying the amplitude of the noise.  
The WMAP team quotes an uncertainty of the noise amplitude of 
0.06\% (at one standard deviation) 
\cite{HinshawBarnes03}\footnote{The information is in appendix B of the published version of this paper. 
It appears to be omitted from all electronic versions of this paper.}.
When including this in our analysis of the mean of the extrema, we find that the WMAP mean statistics
are still qualitatively low, but our results are very sensitive to this value.  
The numerical results are given in figure \ref{amplitudeShiftPHats} and plots of the CDF functions 
are shown in figure \ref{amplitudeShiftFig}. 

Again, with only 110 samples, our statistics are not strong enough to claim a 95\% fluctuation.
In some cases, it is clear that more samples will not help.  
For example, three simulations out of 110 have
more cold spots than the WMAP data, at $N_{side}=512$.  The number of cold spots in the WMAP data will
not likely be a three sigma fluctuation; it will probably be just under two sigma.
For the hot spots, however, all of the simulations have fewer maxima than the WMAP data.  
In this case, more correlated noise simulations would be useful to determine the significance of this result.

Use of the correlated noise has two effects.  It makes the fluctuation in the mean statistic much less significant (if
we also allow a one sigma shift in the noise amplitude).  
Secondly, it indicates that there are too many extrema in the WMAP 
measured CMB, a result not seen in the white noise.  This second effect is not affected by our shifts 
in the amplitude of the noise.

\begin{figure*}
\begin{center}
\begin{tabular}{r|rrrrrrr}
\hline
statistic & $-3\sigma$ & $-2\sigma$ & $-1\sigma$ & $0\sigma$ & $1\sigma$ & $2\sigma$ & $3\sigma$ \\
\hline
Number of Local Extrema max & 0.991 & 1.000 & 0.982 & 0.991 & 0.991 & 0.982 & 0.982\\
Number of Local Extrema min & 1.000 & 1.000 & 1.000 & 1.000 & 1.000 & 1.000 & 1.000\\
Mean max & 0.782 & 0.509 & 0.255 & 0.109 & 0.027 & 0.000 & 0.000\\
Mean min & 0.345 & 0.155 & 0.064 & 0.009 & 0.000 & 0.000 & 0.000\\
Variance max & 0.145 & 0.136 & 0.127 & 0.091 & 0.055 & 0.127 & 0.091\\
Variance min & 0.291 & 0.273 & 0.191 & 0.227 & 0.245 & 0.209 & 0.200\\
Skewness max & 0.455 & 0.509 & 0.464 & 0.473 & 0.373 & 0.455 & 0.409\\
Skewness min & 0.873 & 0.909 & 0.864 & 0.927 & 0.900 & 0.827 & 0.882\\
Kurtosis max & 0.373 & 0.355 & 0.382 & 0.464 & 0.409 & 0.464 & 0.355\\
Kurtosis min & 0.718 & 0.682 & 0.727 & 0.709 & 0.673 & 0.618 & 0.691\\
\hline
\end{tabular}
\end{center}
\caption{
\label{amplitudeShiftPHats}
Number of standard deviations by which the amplitude of the correlated noise was
shifted.  
Hinshaw et al.~\cite{HinshawBarnes03} cite the error in the noise amplitude to be 0.06\%, 
so $-3\sigma$ corresponds to multiplying the amplitude by exactly 0.9982, and $3\sigma$ corresponds
to multiplying by exactly 1.0018, etc.  
This multiplication is carried out after the 
proper scaling of the correlated noise, discussed in section \ref{noiseScale}.
The data for the first two statistics is presented graphically in figure \ref{amplitudeShiftFig}.
}
\end{figure*}

\section{Smoothing One- and Two-point Analysis}

The process of simulation and data reduction is very similar to the previous one, except that we smooth
instead of changing resolution.  The smoothing increases the signal to noise
ratio and therefore reduces our sensitivity to the noise model. 
Also, we are free to choose the smoothing scale without being tied to the discrete pixel sizes of
the HEALPix scheme.
We again describe the simulation and data reduction process.

Our simulation process has only two steps: creating a CMB sky and adding white noise.  Again,
this is an acceptable approximation of the WMAP data if 
our final statistics only depend on the accurate characteristics of this approximation.
To assure this in our data reduction, we again ignore the monopole and dipole moments, as well as the region contaminated
by galactic foregrounds.

\begin{figure*}
\resizebox{12cm}{!}{
	\includegraphics{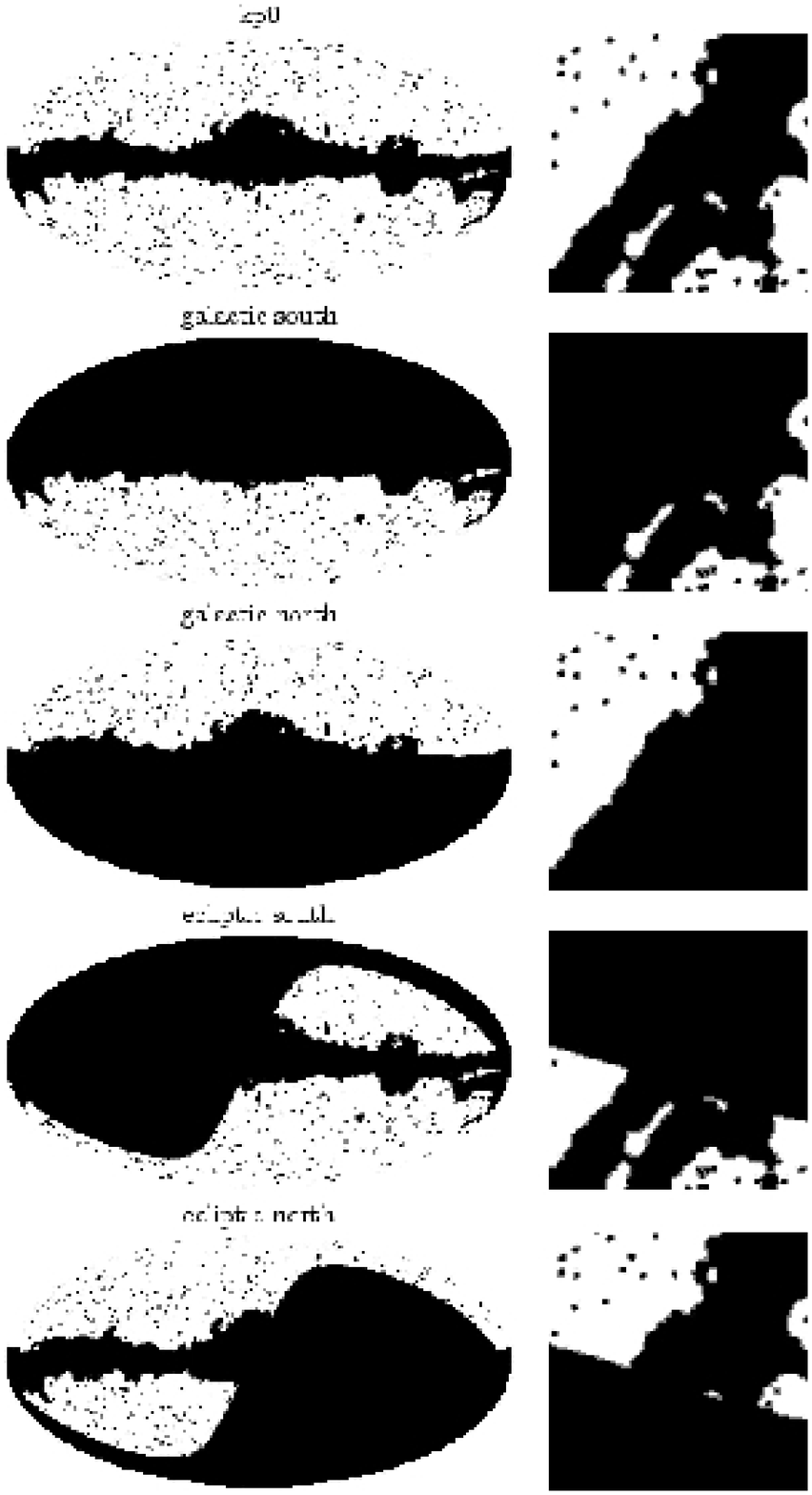}
}
\caption{
\label{twoPointMasks}
These are the 5 masks used for the two-point correlations.  
Left column is mollweide projection of the sky; right column is HEALPix base tile 6, 
where the upper left corner is northernmost.  Tile 6 is directly opposite the galactic center.
The mask is black, the adjusted mask for 50 arcminute FWHM smoothing 
includes the mask and the thin grey region extending the mask. 
}
\end{figure*}

To mask the sky and check for large-scale anisotropies in the statistics, we extend the kp0 mask to 
different hemispheres, as in LW04.  
This yields four other Galactic masks: 
Galactic North and South masks (GN, GS) and Ecliptic North and South masks (EN, ES).
See figure \ref{twoPointMasks}.
When masking a CMB map, we set the temperature fluctuations inside the mask to zero to assure that 
later smoothing does not allow contamination in the masked region to leak out.
To remove dependence on the monopole and dipole, 
we also remove these moments outside of the galactic mask we use.

\begin{figure*}
\begin{center}
\resizebox{12cm}{!}{
	\rotatebox{-90}{
		\includegraphics{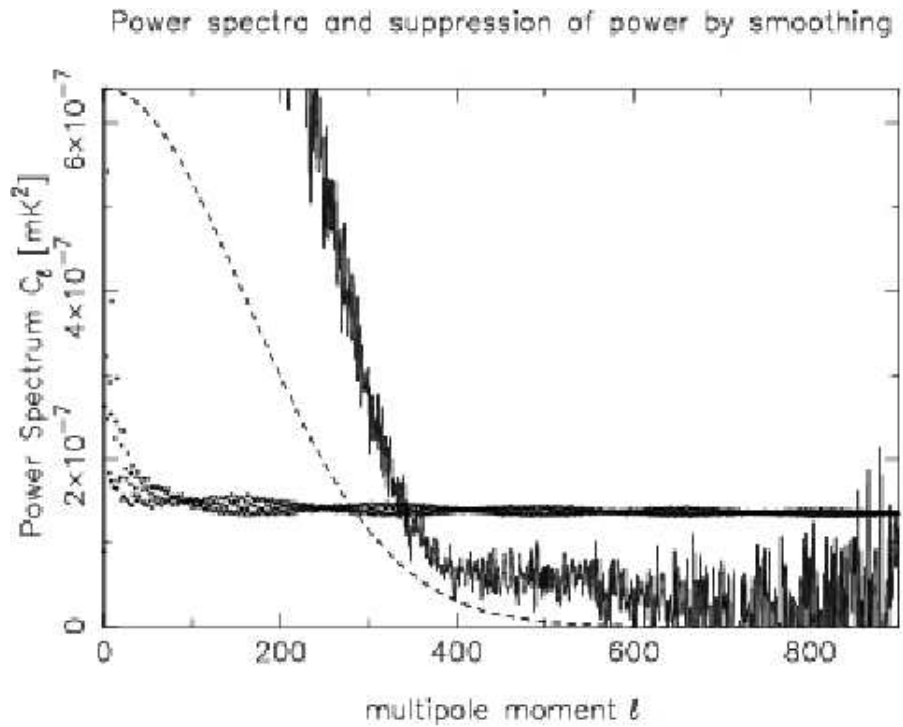}
	}
}
\end{center}
\caption[Power Spectra]{
\label{noisePowerSpectrum}
The solid line is the measured temperature-temperature
power spectrum from the first year WMAP data.  The dots are the average (over
110 simulations) of the W band correlated noise power spectrum.  Note the ringing, which is due
to the pixelization interacting with the scanning strategy.  The dashed line is
the \emph{unitless} suppression of power (due to a 50 arcminute FWHM Gaussian smoothing),
which has been rescaled to fit the height of the graph.  Note that at $\ell = 350$, 
where signal to noise is about 1, smoothing suppresses power by a factor of about 10.}
\end{figure*}

To remove dependence on the small scale structure of the noise, we smooth the sky, 
with either a 50 arcminute or 3 degree Full Width at Half Maximum (FWHM) beam.
To choose the smoothing scale, we arbitrarily decided to suppress the power by a factor
of 10 at the multipole $\ell$ value where the signal to noise has dropped to a ratio of 1.  The 
signal to noise is 1 at about $\ell = 350$, so we choose a Gaussian smoothing
FWHM scale of 50 arcminutes.  As seen in figure \ref{noisePowerSpectrum},
this suppresses the CMB power spectrum by a factor very close to 10 at $\ell=350$. 
We also check our results with 3 degree smoothing scale, where the noise is entirely subdominant.

After the smoothing, we find the maxima and minima.  
However, the smoothed temperature map and therefore these extrema will be affected by
the zeroed pixels inside the mask, so we want to ignore extrema that are significantly affected by the presence of
the mask.
To do this, we create an adjusted mask.  We smooth the original mask (kp0, GS, GN, ES, or EN) 
with the same FWHM Gaussian beam as we will use to smooth
the CMB, and we mask all areas with values less than 0.9.  Recall the convention that unmasked pixels 
have a value of 1 and masked pixels have a value of 0.
When we ignore extrema inside this adjusted mask, 
we ignore most of the extrema which have been significantly affected by being
close to a region of zeroed pixels.
Our value of 0.9 is less conservative than Eriksen et al.~\cite{EriksenBanday04} who use 0.99.
This value affects how strictly we want to ignore mask effects.  
It does not affect the significance of our results, because the simulations and the WMAP data are treated identically.

\begin{figure*}
\resizebox{17cm}{!}{
	\includegraphics{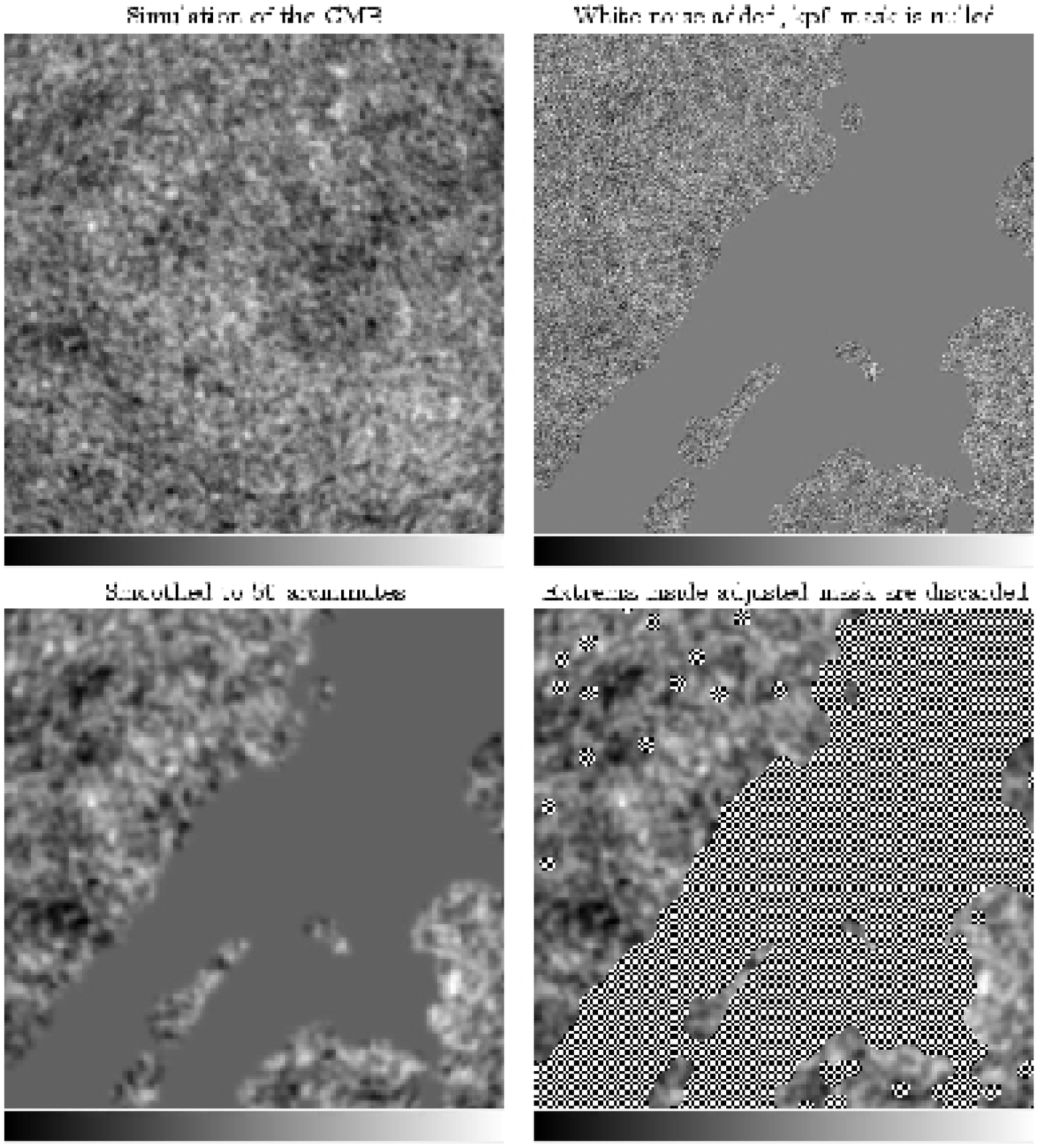}
}
\caption{
\label{twoPointProcess}
These are images of base tile 6 in the HEALPix scheme at various steps in the simulation process.  North is
in the upper left.  Range of color scale varies between some images.
}
\end{figure*}

The process for simulating and and reducing the maps is as follows:
\begin{enumerate}
\item Simulate a map with $N_{side} = 512$, $\ell_{max} = 700$ 
(or $\ell_{max} = 300$, for 3 degree FWHM smoothing), and the WMAP measured power spectrum.
\item Add in white noise, according to the effective number of observations on each pixel.
\item Set the temperature fluctuation to zero inside a galactic mask.
\item Remove the monopole and dipoles outside of that same mask.
\item Smooth with a 50 (or 180) arcminute FWHM Gaussian beam.
\item Find the local extrema.
\item Discard extrema inside the \emph{adjusted} version of the mask in step 2.
\item Calculate statistics on the extrema for further analysis.
\end{enumerate}
A few images of the
various stages of this simulation process are shown in figure \ref{twoPointProcess}.
We compare these simulations to the WMAP cleaned temperature map data, which goes through the same process,
starting at step 3.

\subsection{One-Point Statistics}

We performed 4000 simulations of Gaussian CMB skies.
A 99\% detection would require fewer than 10 of the statistics to be below (or above) the WMAP statistic.
A 95\% detection only requires fewer than 84 of the statistics to be below the WMAP statistic.
See the appendix for details.

\begin{figure*}
\resizebox{17cm}{!}{
	\includegraphics{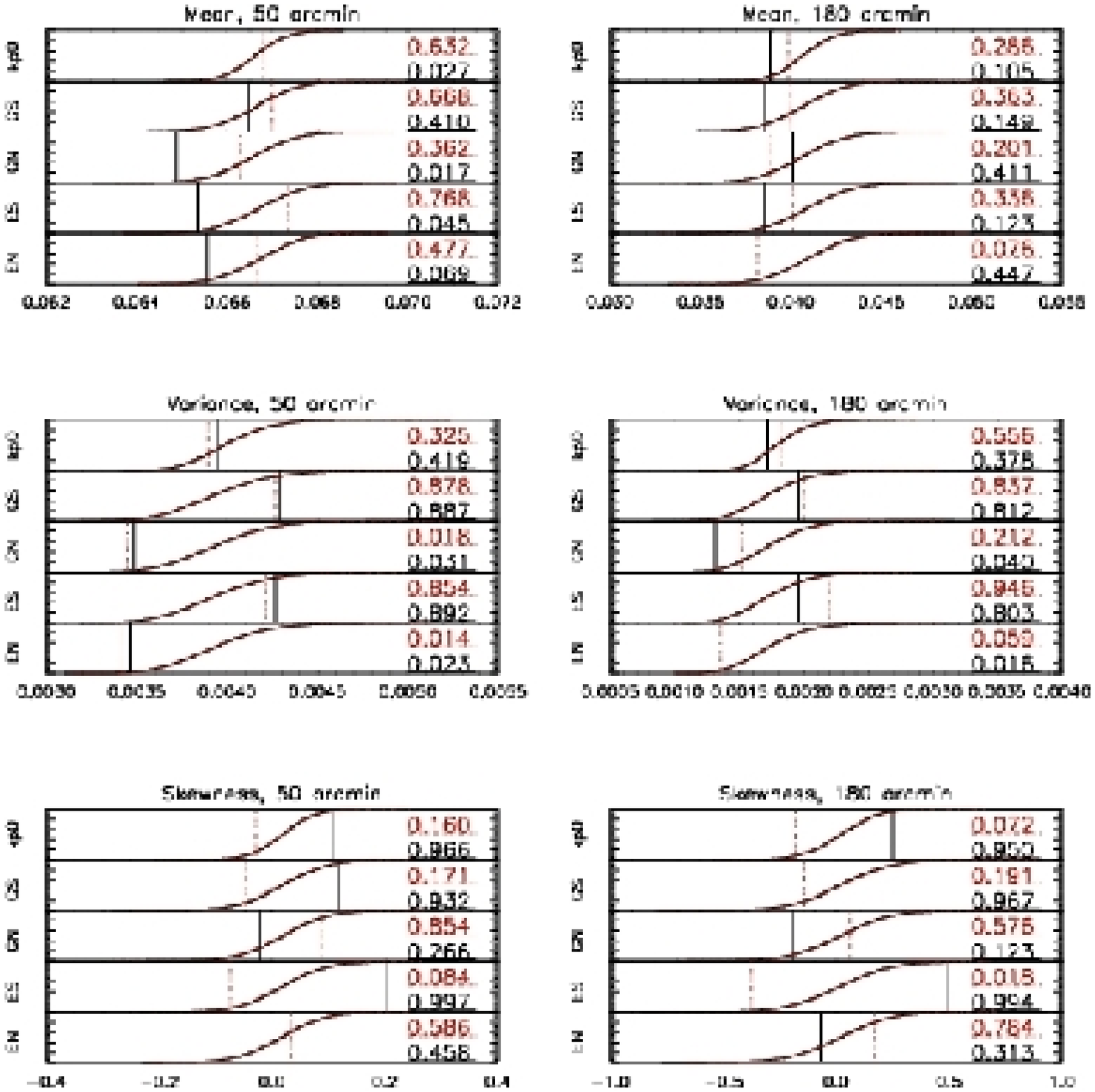}
}
\caption{
\label{onePointSmoothing}
These are cumulative distribution functions of selected one-point statistics for 50 and 180 arcminute smoothing levels.
Hot spot (maxima) statistics are in red and are the upper number in each pair.
Cold spot statistics are in black.
The vertical lines indicate the position of the WMAP statistic.  
Mean statistics are given in $mK$, variance is in $mK^2$, and skewness is dimensionless.
The mean and skewness values of the minima have been multiplied by $-1$ before creating CDFs, for easier
comparison with the maxima.
}
\end{figure*}

Some results for the one-point statistics are shown in figure \ref{onePointSmoothing}.
There are no 99\% detections, but there are several 95\% detections.
For the mean statistic, the hot spots do not seem particularly unusual, but the cold spots are too warm in 
the Galactic North at 50 arcminute smoothing.  
We also have 95\% detections in the Ecliptic North with the hotspots not having enough variance at 50 arcmin smoothing
and the cold spots not having enough variance at 3 degree smoothing.
With respect to skewness, the cold spots have too much negative skewness at both smoothings.  
It is possible that this is caused by a single very cold cold spot, perhaps
that described by \cite{CruzMartinezGonzalez05}. 
The hot spots have too little skewness at the 3 degree smoothing.

We calculate 100 one-point statistics, and 7 of them give 95\% detections,
so one could argue that these detections are not highly significant.
Nevertheless, it is interesting to see that the detections support previous results, such as the lack of
power in the Ecliptic North\cite{EriksenHansen04}.

\subsection{Two-Point Statistics}

\subsubsection{Calculating Two-point Functions}
For the two-point functions, we perform two general analyses on the extrema.
The first is related to the method used by Heavens and Sheth \cite{HeavensSheth99} 
where they look at the 
point-point correlation function of the locations of the maxima above a certain threshold (and minima below
a threshold).  We arbitrarily pick a threshold of $2\sigma$, where $\sigma$ is the standard deviation
of all the temperatures in the map outside the (non-adjusted) mask.

The other analysis is where no threshold is applied to the extrema
and the two-point statistics of the temperature field at
the locations of the extrema are calculated.  It has been proposed that this two-point function
is very close to the two-point function on the full sphere \cite{KashlinskyHernandezMonteagudo01}.  

In both analyses, the statistics we calculate are motivated by the concept of a correlation
function.  We simply find the average number of pairs at a given angular
separation, or the 
average product of spot temperatures at some separation.
We calculate three statistics of this form: 
between maxima and maxima, between minima and minima, and between maxima and minima.

Consider the spot-spot statistics between maxima and minima, for example.
We select all pairs with one hot and one cold spot and find the angles between the spots.  These
angles we bin into 1000 equally spaced bins of angular separation between 0 and $\pi$ radians.
To remove dependence on the number of spots, we normalize the histogram we just made
by dividing by the total number of counts.  This makes the bins sum to 1.
Eriksen et al. \cite{EriksenNovikov04} have already studied the Minkowski functionals
on the CMB, which are related to the number of spots above a threshold, so we did not feel
the need to study it further by including information about the number of spots in our data set.
The normalized histogram contains the correlation information, but it is slightly dependent on
the pixelization and highly dependent on the geometry of the mask we used.

Suppose we wanted to calculate the true underlying correlation function, independent 
of the geometry of the mask.  We do not need to do this in our statistical analysis (and in fact we do not),
since we have exactly the same geometric masking effects included in both the WMAP data
and simulations.  Nonetheless, if we wanted to determine the mask-independent correlation function, 
we would need to know the effects of the mask.
For this purpose, we 
bin the angles between pairs in a random distribution of ``maxima'' and ``minima'' 
for each CMB simulation.  The number of randomly placed extrema is determined by 
the number of extrema found in that simulation.
The underlying correlation function is the excess probability of finding 
pairs at a given angle.  To obtain this for each angular bin, we divide the normalized number of pairs
from the CMB simulation by the average normalized number of random pairs in that bin, and subtract 1.
This gives us a mask independent correlation function for each iteration, which we can use
to visually examine our results.

Calculating the temperature-temperature two-point statistics is very similar to 
calculating the spot-spot statistics.
Instead of counting the number
of angles that fall into a given bin, we find the average product of temperatures
for pairs of spots in that angular bin.  
This can also be turned into a correlation function, if only for visual examination.

\subsubsection{Reducing Two-point Functions to a Single Number}

Now we must reduce a 1000 dimensional discretized two-point statistic $\xi(\theta)$ 
into a single statistic, a single real number.  We treat $\xi$ as a 1000 dimensional vector.
First, we reduce the dimension by ignoring some of the data.  We do this in two ways: 
by re-binning the vector into 40 bins, and by ignoring all but the first 40 of the 1000 bins.
After this, we calculate a $\chi^2$ value for the lower dimensional $\xi$ vector, based on 
the covariance $C$ of those vectors.  We define $\chi^2 \equiv \xi^T C^{-1} \xi$.  
Since we must be sure to treat the WMAP two-point vector in the same way as the simulation vectors,
and we do not want to include it in the calculation of the covariance matrix,
we must use some simulation vectors to define the covariance matrix and calculate our $\chi^2$ statistics
with the rest.
We calculate the
covariance matrix $C$ with the first 1000 vectors $\xi$ and then find where the WMAP data lies in 
the distribution of $\chi^2$ values of the rest of the $\xi$ vectors.  We also visually
check that the distribution of $\chi^2$ values from the vectors used to make the covariance is 
not excessively different from that of the rest of the vectors.  This verifies that we 
are using enough vectors to define $C$.

\begin{figure*}
\resizebox{17cm}{!}{
\includegraphics{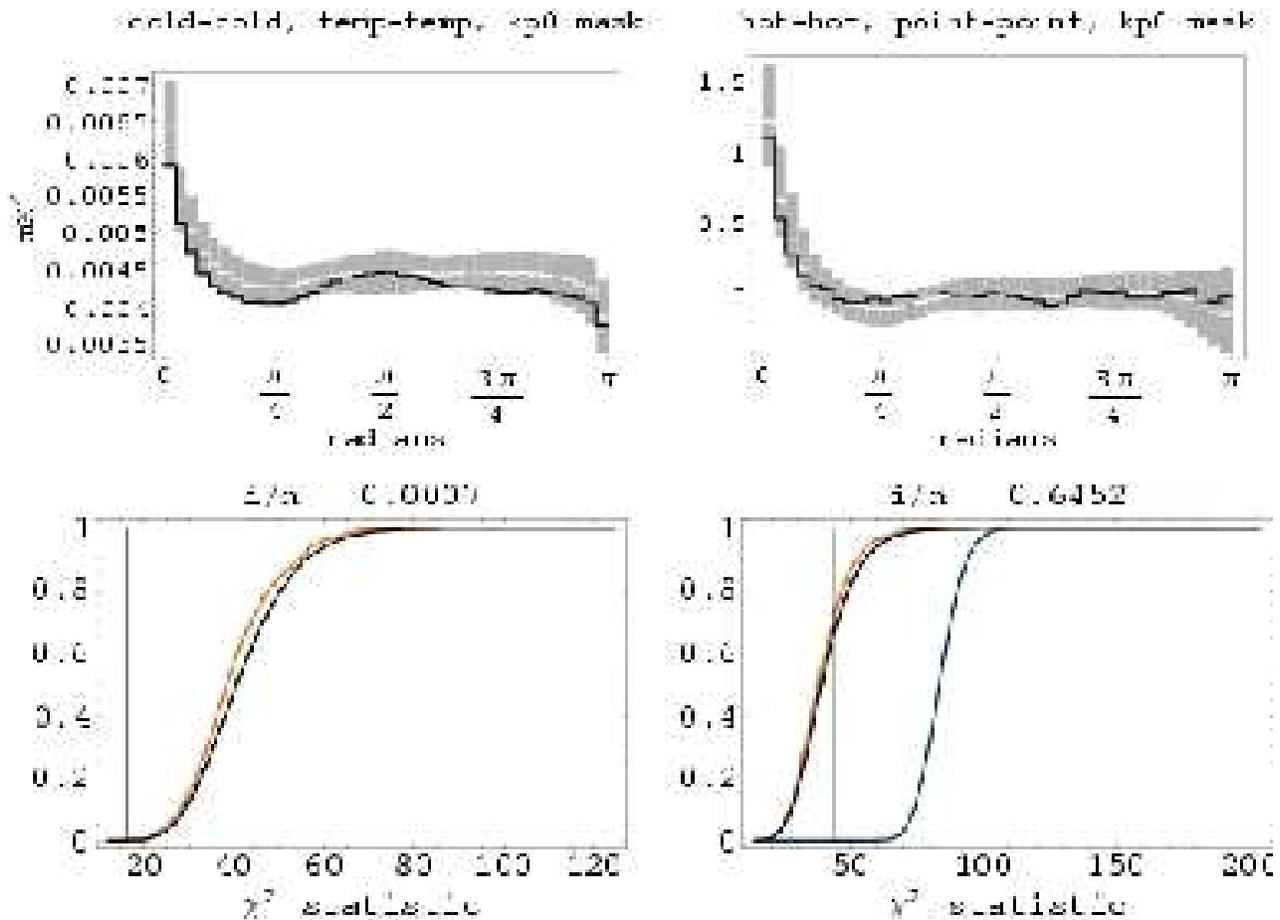}
}
\caption{
Upper left: temperature-temperature correlation, minima, full sky.  Upper right: spot-spot correlation, maxima, full sky. 
The vertical axis is the excess fractional probability density, for finding a pair of points at a given angular separation.
In the correlation functions, the black line is the WMAP data, the white lines are the median simulation values, and the
grey band is a $2\sigma$ error band calculated from the simulations. 
Lower: Cumulative distribution functions for $\chi^2$ statistics.  
They correspond to the upper plots.
}
\end{figure*}

\subsection{Two-point results}

\begin{figure}
\begin{center}
\begin{tabular}{llcccc}
\hline
& & \multicolumn{2}{c}{7.2 degree} & \multicolumn{2}{c}{180 degree} \\ 
& & pt-pt & temp-temp & pt-pt & temp-temp \\ 
\hline
kp0 & min   &  0.14133 &  0.69100 &  0.05033 &  0.00000 \\ 
kp0 & max   &  0.09400 &  0.01000 &  0.38933 &  0.11567 \\ 
kp0 & cross &  0.41833 &  0.18700 &  0.01633 &  0.02067 \\ 
GS  & min   &  0.03567 &  0.85800 &  0.87500 &  0.16300 \\ 
GS  & max   &  0.14633 &  0.10367 &  0.11533 &  0.43467 \\ 
GS  & cross &  0.90200 &  0.59900 &  0.89000 &  0.80567 \\ 
GN  & min   &  0.87400 &  0.28033 &  0.07767 &  0.31033 \\ 
GN  & max   &  0.30600 &  0.21100 &  0.49200 &  0.72667 \\ 
GN  & cross &  0.10233 &  0.06967 &  0.22933 &  0.83233 \\ 
ES  & min   &  0.42667 &  0.77000 &  0.73033 &  0.51933 \\ 
ES  & max   &  0.11500 &  0.16633 &  0.33967 &  0.65267 \\ 
ES  & cross &  0.36467 &  0.38833 &  0.54033 &  0.51900 \\ 
EN  & min   &  0.21833 &  0.20033 &  0.19933 &  0.20533 \\ 
EN  & max   &  0.03100 &  0.07367 &  0.23033 &  0.14067 \\ 
EN  & cross &  0.23700 &  0.18700 &  0.04800 &  0.12700 \\ 
\hline
\end{tabular}
\end{center}
\caption{
\label{50arcminPHats}
These are our estimates $\hat{p}$ of the position of the WMAP two-point statistic
among the simulated statistics.  These results are for 50 arcminute FWHM smoothing, where
1000 iterations went to create the covariance matrix and position
of the WMAP statistic is found among the remaining 3000.
The different rows show results for different masks, as well as the minima-minima, 
maxima-maxima, and minima-maxima statistics.  The columns show the results for the
nearest 7.2 degrees (first 40 bins) of our two-point statistics, as well
as for the full-sky 180 degree two-point statistics.  Columns also show the
different results for the spot-spot and temperature-temperature statistics.
Note the 0 value in the upper right corner.
}
\end{figure}

\begin{figure}
\begin{center}
\begin{tabular}{llcccc}
\hline
& & \multicolumn{2}{c}{7.2 degree} & \multicolumn{2}{c}{180 degree} \\ 
& & pt-pt & temp-temp & pt-pt & temp-temp \\ 
\hline
kp0 & min   &  0.75833 &  0.69500 &  0.83800 &  0.18100 \\ 
kp0 & max   &  0.26333 &  0.16533 &  0.62600 &  0.27000 \\ 
kp0 & cross &  0.95433 &  0.69967 &  0.83500 &  0.14833 \\ 
GS  & min   &  0.40567 &  0.80267 &  0.78467 &  0.42267 \\ 
GS  & max   &  0.98967 &  0.88833 &  0.93700 &  0.52133 \\ 
GS  & cross &  0.00067 &  0.32900 &  0.66933 &  0.75400 \\ 
GN  & min   &  0.34700 &  0.11933 &  0.08233 &  0.03267 \\ 
GN  & max   &  0.59533 &  0.80900 &  0.17967 &  0.60667 \\ 
GN  & cross &  0.65267 &  0.73033 &  0.26500 &  0.80100 \\ 
ES  & min   &  0.10300 &  0.81800 &  0.84533 &  0.51967 \\ 
ES  & max   &  0.62600 &  0.54967 &  0.63733 &  0.84300 \\ 
ES  & cross &  0.82933 &  0.82333 &  0.45500 &  0.71167 \\ 
EN  & min   &  0.67800 &  0.20933 &  0.41333 &  0.17833 \\ 
EN  & max   &  0.75033 &  0.73200 &  0.36067 &  0.33567 \\ 
EN  & cross &  0.69700 &  0.04600 &  0.12133 &  0.17533 \\ 
\hline
\end{tabular}
\caption{
This is the same data as figure \ref{50arcminPHats}, except for 180 arcminute FWHM smoothing.
}
\end{center}
\end{figure}

Our results here are for 4000 simulations.  The first thousand go to define the covariance matrix,
so the WMAP statistic is compared to the statistics for the remaining 3000.  
A 95\% detection requires $0 \le \hat{p} \le 0.02$ or $0.98\le \hat{p} \le 1$,
and a 99\% detection requires  $0 \le \hat{p} \le 0.002$ or $0.998\le \hat{p} \le 1$,
where $\hat{p} = i/n$ is the same as in our previous paper and is also defined in the appendix.

The most interesting result is in the temperature-temperature correlation function for the 
kp0 masked full-sky correlation function.  For 3000 iterations, the WMAP statistic fell
lower than \emph{all} of them.  To better determine the significance of this result,
we ran a set of 20,000 simulations for this particular mask.  Again, the first 1000
went to determine the covariance matrix.  Of the remaining 19,000 simulations, 
13 of their statistics fell lower than the WMAP statistic.  This is extremely close to
a $3\sigma$ detection, it comes to be $2.989\sigma$.

One interesting point about this result is that our $\chi^2$ statistic is too low.  This means that 
instead of fitting the distribution of two-point functions too poorly, it instead fits them too well.
In fact, it fits the covariance matrix describing the two-point functions better than the two-point functions
that went into making that covariance matrix.  This is a very unusual result.

It is also curious that we only see this for the kp0 mask and not for the masks in which we cut out half of the sky.
This suggests that our effect is different from those that led to recent claims of anisotropy in the CMB.

To check to see if this result is an effect of foregrounds, we repeat our simulations for 3000 iterations
in the V-band, just for the kp0 mask.  
We take 1000 simulations for the covariance matrix, and find the position of the WMAP statistic
among the remaining 2000.  
We find that the min-min temp-temp WMAP statistic on the full sky then falls just above 5\% of the simulation statistics.
However, we find that if instead of the min-min statistics we look at the min-max (cross), then we find that
it now sits lower than all 2000 of the simulation statistics.

\section{Conclusions}

In this paper we present frequentist hypothesis tests to check for non-Gaussianity in the CMB using the one and two-point
statistics of hot and cold spots at several angular scales.
We use and advocate a robust statistical test that reduces the probability of 
a false detection of non-Gaussianity to a level commensurate with the significance of the detection.

A \emph{Mathematica} notebook and small user-friendly C program are available 
for determining the significance in this robust hypothesis test where a measurement is compared to Monte-Carlo simulations.
The method and C code are described in the appendix.
The code is available at the web page http://cosmos.astro.uiuc.edu/$\sim$dlarson1/facts/.

While the WMAP hot spots are too cold compared to the white noise simulations, this is no longer
as dramatically true when we use the correlated noise simulations.
The detection drops to below $2\sigma$. 
Instead, we find that there are now too many hot and cold spots in the WMAP data.  We cannot 
give this qualitative statement a significance at or above $2\sigma$ because we only have 110 correlated noise 
simulations.  We also find that the variance continues to be qualitatively low.

While the nature of our result is sensitive to changes in the noise model, the significance is not.
Allowing the amplitude of the white noise to vary by two standard deviations 
puts the CDF squarely around the WMAP measurement.
This means our result did not go away; we may have error either in the amplitude of the noise
or from statistical fluctuations in the CMB, 
but we still need to move something by two standard deviations to make them match.

When we switch to the properly correlated noise samples, regardless of their amplitude 
(within three standard deviations) we definitely get too many cold spots, 
and probably too many hot spots.  This was unexpected, since the white noise 
did not show an unusual number of hot or cold spots.

Our main result comes from our investigation of the two-point statistics.  
We find an anomaly in the full-sky minima-minima temperature-temperature two-point function
using the kp0 mask.  
This is a very large fluctuation, unlikely at the 3 sigma level.  
We observe this anomaly only on the full sky.
This suggests that our effect is distinct from those that led to recent claims of anisotropy in the CMB.

In addition to this 3-sigma result, we also have several 2-sigma results.  For example, we see low variance
of the hot spots at 50 arcminute smoothing, the high skewness of the cold spots at both
50 and 180 arcminute smoothings, and a low ``$\chi^2$'' statistic for 
the point-point function between maxima and minima for
the 180 arcminute smoothing (a 2.7 sigma result).

We have demonstrated that there is a statistically highly significant difference 
between the WMAP data and our Gaussian Monte-Carlo simulations.
This can be interpreted in one of three ways: it is just
a large fluctuation, it is caused by non-Gaussianity, or it caused by some
other unknown foreground or systematic effect that we do not consider in our
model of the WMAP data.  

\acknowledgments{
We benefited from conversations with Olivier Dor\'e and David Spergel.
Some of the results in this paper have been derived using the HEALPix \cite{GorskiHivon99} package.
We acknowledge the use of the Legacy Archive for Microwave Background Data Analysis (LAMBDA). 
Support for LAMBDA is provided by the NASA Office of Space Science.
This work was partially supported by National Computational Science Alliance under 
MCA04N015 and utilized the Xeon Linux Cluster, tungsten.  
BDW gratefully acknowledges a Center for Advanced Study Beckman Fellowship.
}

\newcommand{\cdf}{\mathrm{cdf}}
\newcommand{\rcdf}{\mathrm{rcdf}}
\newcommand{\xj}{ \{x_j\} }
\newcommand{\pI}{\beta}
\newcommand{\pII}{\gamma}
\newcommand{\imax}{i_0}

\appendix

\section{User's Guide and Statistical Methods} 

This appendix contains a brief user's guide for the {\tt facts} program in section \ref{usersGuide} and
a more detailed explanation of our statistical test in section \ref{statTest}.
In section \ref{confidenceInterval} we connect our discussion to the derivation of frequentist confidence intervals
in LW04, and we close with a brief conceptual comment on tests of non-Gaussianity in section \ref{close}.

\subsection{Users Guide for {\tt facts}}
\label{usersGuide}

To aid in the calculation of significance, we provide a publicly available code written in c.  It is named {\tt facts},
which stands for a Frequentist's Ally for the Calculation of Test Significance.
To compile the code on a typical linux system, unzip and untar the file, enter the {\tt facts} directory
which was created, and type {\tt make}.  This will make the executable {\tt facts}.  
The program is small enough that a makefile is not necessary, but I include it for convenience.

The syntax for the command can be retrieved by executing {\tt facts} without 
any command line arguments.  The program takes from three to five arguments.
\begin{alltt}
facts s n i [alpha [beta]]
\end{alltt}
\begin{itemize}
\item {\tt s}: this value is either 1 or 2 for a single or double sided test.
\item {\tt n}: this is the number of simulated statistics calculated.
\item {\tt i}: this is the number of simulated statistics that fell below the test statistic.
If {\tt i} is a negative number, the code prints out information on what values of {\tt i} will reject the hypothesis.
\item {\tt alpha}: this is a number between 0 and 1 that parameterizes our hypothesis.  
For a single sided test, the hypothesis is $p\in(\alpha, 1]$.  
For a double sided test, the hypothesis is $p \in (\alpha/2, 1-\alpha/2)$.
If not specified, the default value is $\alpha = 0.05$.
\item {\tt beta}: this is a number between 0 and 1 that gives an upper limit on the probability of a type I error 
(rejection of the hypothesis when the hypothesis is actually true).  
If not specified, it takes the value of $\alpha$.
\end{itemize}

The code takes the values of {\tt s}, {\tt n}, $\alpha$, and $\beta$ and constructs a test.  If {\tt i}
satisfies that test then the hypothesis is accepted, otherwise it is rejected.  The code prints out information
whether {\tt i} satisfies the test, and what the minimum values of $\alpha$ and $\beta$ are for
which {\tt i} will satisfy the test.

A few examples follow.  The next command determines whether a two-sided test with a $0.02$ maximum
probability of a type I error will accept the hypothesis that $p\in(0.005,0.995)$ when only 17 out of 10000
statistics fell below the test statistic.
\begin{verbatim}
facts 2 10000 17 0.01 0.02 
\end{verbatim}
The command below determines if 5 out of 1000 statistics falling below the test statistic is few enough
to reject the hypothesis $p\in(0.025,0.975)$ when the test has a 5\% maximum chance of a type I error.
\begin{verbatim}
facts 2 1000 5   
\end{verbatim}

While this program is reasonably fast, it is not optimized for pathological requests.  It is slower
for larger values of {\tt n}, {\tt i}, $\alpha$, and $\beta$.

\subsection{Statistical Testing}
\label{statTest}

\subsubsection{The Problem}

The statistical analysis we discuss in this paper can be reduced to the following
problem.  We are given a few thousand ($=n$) random numbers $\xj$, $j=1\ldots n$ which have come from 
some random number generator (distribution), with some probability density function (PDF) $f(x)$.
We are also given a single number $x_0$ and asked to determine
if we have any reason to believe, statistically, that it may not have come from that same 
random number generator.

\subsubsection{The Solution}

Because this is a statistical problem, we cannot do what we naturally want to do: prove that $x_0$ 
was or was not chosen from the same distribution as the $\xj$.  Instead we must settle for a weaker
statement about how large a fluctuation $x_0$ is, if it were drawn from the distribution of the $\xj$.
This appendix describes how to make statistical statements about the size of this fluctuation.

We begin by assuming that the single number $x_0$ did come from the random number
generator that produced the $\xj$. 
Given this, we attempt to determine how large a statistical fluctuation $x_0$ is.
There are several ways to do this. 
We will use a completely standard (frequentist) hypothesis test.  
Our hypothesis concerns a parameter, $p$, which describes how large the fluctuation is.  
We perform a test on a random variable, $i$, whose distribution is affected by $p$, 
and, based on the results of that test, decide whether or not to accept the hypothesis.

This method is useful if the pdf is sufficiently complex that 
it is not practical to evaluate it analytically.

\subsubsection{The Random Variable and Its Distribution}

The method we use requires knowing only how many of the random numbers $x_j$ fell below $x_0$.
Let there be $n$ random numbers and let $i$ of them fall below $x_0$.
Then $i$ is our random variable, chosen from the binomial distribution $P(i|p,n)$, where
\begin{equation}
P(i | p, n) = \frac{n!}{i!(n-i)!} p^i p^{n-i}
\end{equation}
and where $p$ gives the position of $x_0$ in the PDF $f(x)$:
\begin{equation}
p \equiv \int_{-\infty}^{x_0} f(x)\;dx
\end{equation}
We can estimate $p$ with $\hat{p} = i/n$, which is a maximum likelihood and unbiased estimator.

\subsubsection{The Hypothesis}

Our hypothesis concerns the value of $p$.  
As previously stated, we would prefer to pick ``$x_0$ was chosen from the same distribution as the $\xj$ values'' 
as our hypothesis.
Since the negation of this hypothesis is extremely difficult to work with, we choose a simpler hypothesis, about $p$:
\begin{equation}
H_0: \qquad p \in (\alpha/2, 1-\alpha/2) 
\end{equation}
where $\alpha$ is much less than $1$.
Here we use a double sided hypothesis for our test; 
the case of a single-sided test is discussed in section \ref{oneSidedSection}.
To a frequentist, the hypothesis $H_0$ is either true or false, so it either has a probability of 1 or 0. 
The statistical test described in this appendix will then be useful to decide whether to accept or reject the hypothesis. 
The frequentist statement
is that ``$H_0$ is true in a fraction $1-\alpha$ of all possible Gaussian Universes''.

It is certainly true that we could have chosen our hypothesis to be anything of the
form $p\in S$ where the set $S \subset [0,1]$ has total length (or measure) $1-\alpha$.
We choose $S = (\alpha/2, 1-\alpha/2)$ because of our natural inclination to think that values of $p$ far out in the
tails of the distribution are unusual.
Also, if we had reason to believe that $x_0$ were drawn from a distribution whose mean was many standard deviations
away from the distribution of the $\xj$, 
our hypothesis would be a powerful test of whether $x_0$ was drawn from the same distribution
as the $\xj$.

\subsubsection{The Test and Types of Error}

Our test must be of the form where we accept $H_0$ for certain values of $i$, $i\in I$,
and reject it for all others, $i\in\overline{I}$.  
Here, $I$ and $\overline{I}$ are disjoint and $I \cup \overline{I} = \{0,1,2,\ldots,n\}$.
With a statistical test of this form, one is interested in the errors of type I (rejection of a true hypothesis) 
and type II (acceptance of false hypothesis).  I will call their probabilities $\pI$ and $\pII$, respectively.
\begin{center}
\begin{tabular}{l|c|c|}
& test accepts $H_0$, & test rejects $H_0$, \\ 
& $i\in I$ & $i \in \overline{I}$ \\ 
\hline
$H_0$ is true & $1-\pI$ & $\pI$ \\
\hline
$H_0$ is false & $\pII$ & $1-\pII$ \\
\hline
\end{tabular}
\end{center}

Explicit calculation of these probabilities requires us to specify a test.  
Our test is similar in form to that of our hypothesis.
We accept $H_0$ whenever 
\begin{equation}
i \in \{\imax+1, \imax+2\ldots, n-\imax-1 \}
\end{equation}
for some specified value of $\imax$.
Since calculation of $\pI$ and $\pII$ requires knowledge of $n$, $p$, and the test used, 
$\pI$ and $\pII$ are functions of these three 
values.
\begin{equation}
\pI = \pI(\imax, p, n)\qquad \pII = \pII(\imax, p, n)
\end{equation}

It is desirable to adjust the test (the value of $\imax$) 
so that $\pI$ and $\pII$ are as small as possible for all values of $p$.  
Unfortunately, these cannot both be made small.
The probability of the test accepting $H_0$ must vary continuously as $p$ varies continuously 
from the region from where $H_0$ is true to where it is false. 
This means that $\pI(\imax, p, n) = 1 - \pII(\imax, p, n)$ at $p=\alpha/2$ and $p=1-\alpha/2$.
We must decide which we want to be small.  
For this paper, we choose to make $\pI$ small.  
This all but eliminates the possibility of
rejecting $H_0$ when it is actually true.  
As a trade-off, we have the problem of accepting  $H_0$ when it may be false.

We can construct an explicit expression for $\pI$:
\begin{equation}
\label{pIequation}
\pI(\imax, p, n) = \sum_{i=0}^{\imax} P(i | p, n) + \sum_{i=n-\imax}^n P(i | p, n)
\end{equation}
We want to limit $\pI(\imax, p, n)$ for any value of $p$ satisfying $H_0$, 
which means we want to limit $\pI(\imax, \alpha/2, n)$.  There are two approaches to this: 
to fix the hypothesis ($\alpha$) and change the test ($\imax$), or to fix the test and change the hypothesis.
Suppose we are given the hypothesis and want to determine a test that keeps 
$\pI$ below some bound.  
This requires one to progressively increase $\imax$ while checking that $\pI(\imax, \alpha/2, n)$ 
remains below the desired bound.
Alternatively, if we have done our experiment and want to determine the highest possible significance we can
assign to it, then we know $\imax$ and want to find $\alpha$.  Specifically, we set $\imax$ to our measured value of $i$, 
and find the smallest value of $\alpha$ such that $\pI$ is still below some desired bound.  For simplicity, we
could set $\alpha = \pI$, since we want both values to be low for a highly significant result, and numerically solve
for $\alpha$:
\begin{equation}
\pI(\imax, \alpha/2, n) = \alpha
\end{equation}
We can then claim that our test has rejected the hypothesis, 
and has a probability of a type I error (rejecting a true hypothesis) of $\alpha = \pI$.

For computational purposes, it is useful to note that
one of the sums in equation \ref{pIequation} will contribute very little to the value of $\pI$. 
Let us consider the contribution of the second sum, when it is the smaller of the two.
Its maximum value (while still being smaller) occurs when $p = 1/2$.  We will also estimate the 
integer $\imax+1 \approx n\alpha/2$. The value of $\imax+1$ will typically be lower than this.
\begin{eqnarray}
\label{smallNumber}
\sum_{i=n-\imax}^n P(i | \alpha/2, n) & < & (\imax+1) P(n-\imax | \alpha/2,n) \nonumber \\
& < & (\imax+1) n^{\imax+1} (1/2)^{n} \nonumber\\
& \lesssim & (n \alpha/2) n^{ n\alpha/2} (1/2)^{n} \nonumber\\
& \lesssim & \exp\{ \ln(n \alpha/2) +(n\alpha/2) \ln(n) \nonumber \\
&& + n \ln (1/2) \}
\end{eqnarray}
We want this to remain small.
To assure that the $\alpha n\ln n$ term does not become larger than the $n\ln(1/2)$ term as $n$ increases, 
we must have $\alpha \le k/\ln(n)$ for some $k$.  
Now we can check the contribution of the second sum (equation \ref{pIequation}) 
to $\pI$ for some reasonable bounds:
\begin{equation}
\label{sumAssumptions}
\alpha \le 1/\ln n \qquad n \ge 100 \qquad \imax \le \alpha n/2.
\end{equation}
We find the value of the second sum in equation \ref{pIequation} 
to be well below $10^{-40}$ whenever these conditions are satisfied. 
Since the probabilities we test are all much higher than $10^{-40}$, 
it is safe to ignore the second sum in equation \ref{pIequation} 
when $\hat{p} \ll 1/2$.
By symmetry, it is safe to ignore the first sum when $\hat{p} \gg 1/2$.  
The {\tt facts} code makes use of this information by ignoring the second sum
and mapping its input to a problem where only the first sum is important.

\subsubsection{The Case of a One-Sided Test}
\label{oneSidedSection}

It may be useful for other applications to do a one-sided analysis, for example if one will only consider
a statistic $x_0$ to be unusual if it is lower than most simulated statistics.  Our work can be
repeated for that case:
\begin{equation}
H_0:\qquad p\in(\alpha,1]
\end{equation}
We accept the hypothesis when $i > \imax$.
\begin{equation}
\pI(\imax, p, n) = \sum_{i=0}^{\imax} P(i | p, n)
\end{equation}
To specify a test, we require $\imax$ to be small enough that the maximum value of $\pI$, 
which is $\pI(\imax, \alpha, n)$, is below some bound.  
Alternatively, to find the significance of previously obtained results, 
we set $\imax$ to be our measured value of $i$, set $\alpha = \pI$, and solve
\begin{equation}
\pI(\imax, \alpha, n) = \alpha
\end{equation}
to find the maximum significance of our test.
As in the double sided case, we can claim that our test rejects the hypothesis $H_0$, and our test has
a probability of a type I error (rejecting a true hypothesis) of $\alpha = \pI$.
The analysis for a single sided confidence interval $H_0: p\in[0,1-\alpha)$ can be mapped by symmetry to
the above problem.

\begin{figure*}
\begin{center}
\resizebox{18cm}{!}{
\includegraphics{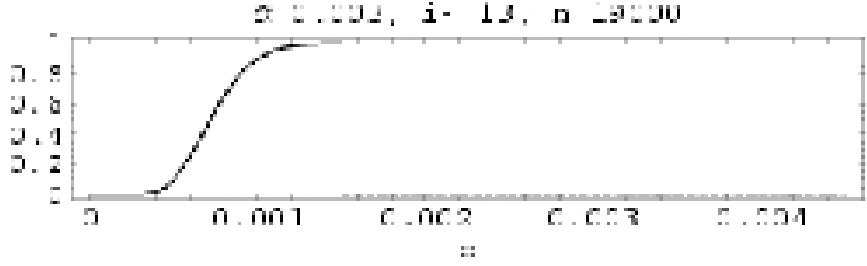}
}
\end{center}
\caption{
This figure plots $\pII$, the probability of a false acceptance of the hypothesis (solid), 
and $\pI$, the probability of a false rejection (dashed), 
as a function of $p$, for $\alpha = 0.003$, $\imax=13$, $n=19000$.  In order to keep $\pI$ below $\alpha$,
we find that $\pII$ becomes quite large for some values of $p$.  
In order to avoid false rejections of the hypothesis, we must allow false acceptances sometimes.
The region in which $\pII$ is large becomes smaller as $n$ increases. 
}
\end{figure*}

\subsection{Connecting to Frequentist Confidence Intervals}
\label{confidenceInterval}

For an understanding of our statistical test that is sufficient to do calculations,
the previous section is enough, and the practical reader can stop here.  For the 
enthusiastic reader, we describe in this section how our test can
also be derived using frequentist confidence intervals.  
This connects the preceding discussion to our alternative derivation for the calculations
in LW04.

We use the same formalism in the previous section.  There are $n$ simulated
statistics, $\xj$.  Exactly $i$ of these fall below the test statistic $x_0$.
The true probability of another simulated statistic falling below the test statistic
is $p$, which can be estimated by $\hat{p} = i/n$. 

We construct an interval $[p^-, p^+]$ such that the true value of $p$ will be inside 
the interval at least a fraction $1-\beta$ of the time.  
Note that this doesn't mean we 
think the probability of $p\in [p^-,p^+]$ is at least $1-\beta$ for the specific interval we construct, since
$p$ is not a random variable.
For some given interval, $[p^-,p^+]$, either $P(p\in [p^-,p^+]) = 0$ or  $P(p\in [p^-,p^+]) = 1$, and we do not
know which is correct.  Instead it is helpful to think about many
sets of numbers $\xj$ and the same $x_0$.  Each set will have $n$ numbers, where $i_k$ of
them fall below $x_0$.  We construct $\hat{p}_k = i_k / n$ as before.  For each set we also
construct the interval $[p^-_k,p^+_k]$.  When we say the probability $P(p\in [p^-,p^+]) \ge 1-\beta$,
this is to be interpreted with $p^-$ and $p^+$ as the random variables, since they are both
functions of the random variable $i_k$.  It is a statement
about whether the interval falls around the true value of $p$ and not about whether $p$ falls
in the interval.

The procedure for constructing this interval $[p^-, p^+]$ is detailed in chapter 20 of 
Kendall \& Stuart \cite{KendallStuart73}.
We state a few relevant results here.

We define a cumulative distribution function over the index $i$ and a reverse cumulative distribution function
as follows:
\begin{equation}
\cdf(i,p,n) = \sum_{j=0}^{i} P(j|p,n)
\end{equation}
\begin{equation}
\rcdf(i,p,n) = \sum_{j=i}^{n} P(j|p,n)
\end{equation}
From here, we can pick a significance $1-\beta$ for our test and then define 
$p^-=p^-(i,n,\beta)$ and $p^+=p^+(i,n,\beta)$ by
\begin{eqnarray}
\cdf(i,p^+,n) & = & \beta/2 \\
\rcdf(i,p^-,n) & = & \beta/2 
\end{eqnarray}
Note that here we have constructed a double sided confidence interval.  If we wanted a single sided
interval, for example $p\in[0,p^+]$ with probability $\ge 1-\beta$, then we would have 
$p^+=p^+(i,n,\beta)$ defined by
\begin{equation}
\cdf(i,p^+,n)  =  \beta
\end{equation}
We will not prove this in mathematical detail, but we will restate an argument given by \cite{KendallStuart73} for
the double sided confidence intervals.

\begin{figure}
\begin{center}
\resizebox{9cm}{!}{
\includegraphics{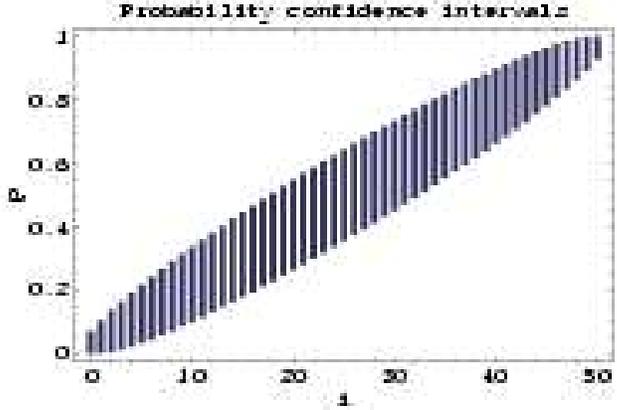}
}
\end{center}
\caption{
\label{confidenceIntervals}
Double sided confidence intervals, $\alpha=0.05$, for all values of $i$ from 0 to $n=50$.
}
\end{figure}

As \cite{KendallStuart73} describes, figure \ref{confidenceIntervals} 
can be constructed horizontally and read off vertically.
We construct it by fixing $p$ and varying $i$ (as we sum cdf() and rcdf()) 
while we read off the numbers $p^-$ and $p^+$ at fixed $i$.

Consider a single value of $p$.  Then look across at the values of $i$ that are marked in blue at that $p$ value. 
By construction, they have probabilities that sum to $ \ge 1-\beta$.  Hence for any $p$ value, the 
probability of being in the blue band is $\ge 1-\beta$.  But one is in the blue band if and only if
$p\in [p^-,p^+]$.  Hence the interval $[p^-,p^+]$ will fall around the correct value of $p$ with probability
$\ge 1-\beta$.    I'll rephrase it once again:
At any specific $p$ value, values of $i$ that cause the interval $[p^-,p^+]$ to surround $p$
will be chosen at least $1-\beta$ of the time.

This frequentist confidence interval can now be used to construct a test of our hypothesis, 
$H_0$: $p\in(\alpha/2,1-\alpha/2)$.
We reject the hypothesis if and only if our interval  $[p^-,p^+]$ is entirely  outside of the interval 
$(\alpha/2,1-\alpha/2)$.
In LW04, we required a double sided confidence interval $[p^-,p^+]$ to be outside of $(\alpha/2,1-\alpha/2)$.  This was too
conservative, since we only needed a single sided confidence interval to be outside of $(\alpha/2,1-\alpha/2)$.
That is, 
for low  values of $\hat{p}$, we reject $H_0$ if $[0 ,p^+]$ is completely below the interval ($\alpha/2,1-\alpha/2)$, 
which means $p^+ < \alpha/2$. 
For high values of $\hat{p}$, we reject $H_0$ if $[p^- ,1]$ is completely above the interval ($\alpha/2,1-\alpha/2)$, 
which means $1-\alpha/2 < p^-$.

This is equivalent to the test previously discussed, as long as the reasonable assumptions in equation \ref{sumAssumptions}
hold so that one of the sums in equation \ref{pIequation} can be dropped.

\subsection{The Next Step}
\label{close}

We can now quantify how large a statistical fluctuation the number $x_0$ is 
among the random numbers $\{x_j\}$.  
Given the number of statistics that fell below the test statistic,
we can test the hypothesis that $x_0$ is not in the tails (which have total probability $\alpha$) of the distribution.
If our test says that $x_0$ is indeed far out in the tails of the distribution, then we can decide 
not to believe that $x_0$ came from the same random number generator as the $\{x_j\}$.
This is in fact the route taken in our paper.

A general caveat of any blind search for anomalies is that one cannot distinguish between
a large statistical fluctuation and a genuinely different parent distribution.
This is often casually referred to as the ``non-dog'' problem of non-Gaussianity searches,
since the class of things which are not dogs is as large and as difficult to deal with
as the class of CMB models which are not Gaussian.  The difficulty lies in defining 
something by negation.

To illustrate this idea, 
suppose we find a detection for some very small value of $\alpha$.  
Claiming that $x_0$ was not produced by the original random number generator (RNG)
is a dangerous thing to do without specifying exactly what random number generator did
produce $x_0$.
Consider, for example, that the $\{x_j\}$ are Gaussian distributed with zero mean
and unit variance, and $x_0 = 10$.
If the only alternative RNG is one which also has zero mean, but has a variance of
$10^{-100}$, then what we had originally claimed as evidence against the original RNG
would now be considered strong evidence for it.

The ambiguity made apparent in this example holds 
true generally for any blind test of non-Gaussianity and is not just a 
feature of our work.
Consequently, we must either come up with an alternative model to Gaussianity, or 
leave the interpretation to the reader.

\bibliographystyle{unsrt}
\bibliography{annotatedOut}

\end{document}